
\input topp.tex
\PHYSREV
\tolerance 2000
\nopubblock

\titlepage
\title{\bf Phase Ordering Dynamics in the Continuum q-state Clock Model}
\author{Fong Liu
        and Gene F. Mazenko}
\address{James Franck Institute and Department of Physics\break
         University of Chicago, 5640 South Ellis Avenue\break
         Chicago, Illinois 60637}

\vfil
\abstract
\noindent

The order parameter correlation function of
the nonconserved, continuum $q$-state clock model is evaluated
in the asymptotic scaling limit, during the phase ordering
process after a temperature quench.
The short distance behavior of the order parameter scaling function
exhibits explicit crossover from that characteristic of the Ising
universality class to that of the $O(2)$ model.

\bigskip\vfil
\endpage

The phase ordering process of a system quenched from a high temperature
disordered state into the ordered phase exhibit {\it universal}
scaling behavior in the asymptotic late
stages\rlap.\Ref\review{For reviews see, J.D. Gunton, M. San Miguel,
 and P.S. Sahni in {\it Phase Transitions and Critical Phenomena},
edited by C. Domb and J.L. Lebowitz (Academic, New York, 1983),
Vol. 8; H. Furukawa, \journal Adv. Phys. &34&703(85);
 K. Binder, \journal Rep. Prog. Phys. &50&783(87).}
The  characteristic scaling length for pure systems grows
algebraicly in time, $L(t)\sim t^n$ where $n$ is called the growth exponent.
In addition, the order-parameter correlation function satisfies
similarity scaling
 $$ C\rt= \psi_0^2 {\Cal F}(\vecr/L(t)) \eqno\eq $$
where $\psi_0$ is the ordered magnitude of the order parameter.
The set of factors that characterize a universality class,
 and in particular the
functional form of $\Cal F$, is known to
include at least the dimensionality of space, whether the order parameter
is conserved, and the degeneracy of the ground states.

We shall be concerned here with this last factor for
a system with nonconserved order parameter.  The importance of
the degeneracy of ground states lies in the fact that the symmetry
property of the ground state uniquely determines the types of
topological defect structures which control the late stage ordering
in unstable systems. Recent studies of systems with continuous
symmetries\REFS\nigel{
M. Mondello and N. Goldenfeld, \journal Phys. Rev. A. &42&5865(90);
\journal ibid. &45&657(92).}\REFSCON\bray{
A.J. Bray and S. Puri, \journal Phys. Rev. Lett. &67&2670(91).}\REFSCON\fong{
Fong Liu and G.F. Mazenko, \journal Phys. Rev. B. &45&6989(92);
and ``Defect-defect correlations in the dynamics of first-order phase
transitions'', {\sl Phys. Rev. B.}, in press.}\REFSCON\toyoki{H. Toyoki,
\journal Phys. Rev. B.
&45&1965(92).}\REFSCON\braytwo{A.J. Bray and K. Humayun, \journal
J. Phys A. &25&2191(92).}\refsend (with infinitely
degenerate ground states) shows a variation in the
behavior of the scattering structure factor $\Phi(k)$, the Fourier
transform of $C\rt$, from the
usual scalar Ising type systems. Namely, for large wavenumbers,
Porod's law, $\Phi(k)\sim k^{-(d+1)}$, for a scalar system is
replaced by $\Phi(k)\sim k^{-(d+n)}$ for the general $O(n)$ model.

In this paper we investigate the phase ordering process for the
continuum version of the $q$-state clock model\Ref\potts{
R.B. Potts, \journal Proc. Cambridge Phil. Soc. &48&106(52).}
 ( also referred to as the planar Potts model or the vector
Potts model).
This model provides an interesting {\it intermediate} case
between the scalar Ising-type dynamics (where $q=2$) and the model A
dynamics with a complex order parameter (where $q=\infty$).
Unlike the latter two
limits, where the topological defects are interfaces and
vortices respectively, the defects in $q$-state clock model involve
both interfaces and vortices reflecting the $q$-fold degeneracy of
the ground states\rlap.\Ref\kawasaki{
K. Kawasaki, \journal Phys. Rev. A. &31&3880(85).}
This model clearly is relevant to understanding
the disorder-order transition in some alloys\rlap,\Ref\nagler{
S.E. Nagler, \etal, \journal Phys. Rev. Lett. &61&718(88);
S. Katano, \etal, \journal Phys. Rev. B. &38&2659(88).}
  as well as in the
evolution of cellular structures ubiquitous in nature\rlap.\Ref\cahn{
For a recent review, see H.V. Atkinson, \journal
Acta Metall. &36&469(88).}

Many numerical studies have been devoted to the investigation of
phase ordering of clock models\Ref\kaskiclock{
K. Kaski and J.D. Gunton, \journal Phys. Rev. B. &28&5371(83).}
and the standard Potts
models\rlap.\Ref\anderson{
See M.P. Anderson, G.S. Grest, and D.J. Srolovitz, \journal
Phil. Mag. B. &59&293(89); S. Kumar, J.D. Gunton, and
K.K. Kaski, \journal Phys. Rev. B. &35&8517(87), and references therein.}
While it is by now well established that $L(t)\sim t^{1/2}$ in
both models, there is considerable uncertainty in our understanding
of the correlation
functions. Numerical work generally shows independence of the
correlation functions on the degeneracy $q$ of ground states.
On the theoretical front, Kawasaki\refmark{\kawasaki} extended the method of
Ohta-Jasnow-Kawasaki\Ref\ojk{T. Ohta, D. Jasnow, and K. Kawasaki,
\journal Phys. Rev. Lett. &49&1223(82).}
 to the clock model, but did not determine the explicit
form of the correlation function. In this paper, we
shall derive explicitly the short distance behavior of the order-parameter
correlation function in the asymptotic scaling regime. The auxiliary
field method we adopt here has been proven useful in previous
growth kinetics studies\rlap.\REFS\gene{G.F. Mazenko, \journal
Phys. Rev. Lett. &63&1605(89); \journal Phys. Rev. B.
&42&4487(90).}\REFSCON\fongtwo{
Fong Liu and G.F. Mazenko, \journal Phys. Rev. B. &44&9185(91);
  \journal ibid. &45&4656(92).}\refsend

The dimensionless free energy for the continuum $q$-state clock
model\Ref\enomoto{Y. Enomoto, T. Aokage, and K. Isibasi,
\journal J. Phys. Cond. Matt. &4&L133(92).}
($q \geq 2$) can be written as
$$ F=\int d\vecr \left( |\nabla\psi|^2 -|\psi|^2+ \frac{1}{2}
       |\psi|^4 +\frac{v}{q}(\psi^q +\psi^{*q}-2|\psi|^q)\right)  \eqno\eq $$
where $\psi\rt=|\psi\rt|e^{i\phi\rt}$ is the complex order parameter,
and $v>0$ is a constant. By assuming the standard relaxational Langevin
dynamics for the evolution of the order parameter
we obtain the equation of motion
$$ \frac{\pa \psi}{\pa t}= \nabla^2 \psi +\psi(1-|\psi|^2)
     -v (\psi^{*q}-|\psi|^q)/\psi^*   \eqn\motion $$
where the noise term has been neglected for a zero temperature
quench. The $q$ degenerate ground states are then given simply by
$|\psi|=1$, $\phi_j=\frac{2\pi}{q}\left(j-\frac{1}{2}\right)$ with
$j=1,2,...,q$, corresponding to $q$ equivalent directions of a clock.

The phase ordering process of the clock model involves the motion
and annihilation of both interfaces and vortices. Interfaces
separate distinct ground states, while vortices are places
where more than two interfaces meet. At the late stage of the ordering,
it is reasonable to assume that only the topologically stable
defect configurations with the lowest energy will be dominant.
Such a defect configuration is simply a vortex where all $q$-distinct
interfaces meet, so that the phase of the order parameter changes
by $2\pi$ going around the vortex. The order parameter field
around an isolated defect can be described
by the inhomogeneous solution of the equation of motion \motion.
The {\it far field} solution for a vortex centered at $\vecr=0$
can easily be found to be of the form
$$   \psi_{far}(\vecr) \sim \exp \frac{2\pi i}{q} \left[
     \frac{\theta(\vecr)}{2\pi/q}\right]  \eqn\farfield $$
where $\theta$ is the polar angle of $\vecr$ and
we introduce the notation that $[a]$ is the largest integer
not exceeding $a$. To examine the solution near an interface but far away from
the vortex core, we  extract from \motion\ the effective equation of motion for
the variation of phase degree of freedom only and obtain
$$  \phi''(z) +v \sin \left(q\phi(z)\right)=0  \eqn\gordon $$
where $z$ is the effective coordinate perpendicular to the domain wall.
Equation\gordon\ is of the sine-Gordon form and has the soliton-type
solution
$$ \phi(z)=\frac{4}{q}\tan^{-1}\exp (\sqrt{qv}z) +\frac{2\pi j}{q}
 \eqn\solgordon $$
 near the $j$th domain-wall.

 In the context of growth kinetics the key question is the structure of the
scaling behavior at long time and large distances. As usual we assume that
in this asymptotic scaling regime, the order parameter structures can be
associated with an auxiliary Gaussian field\rlap.\refmark{\fong}
In our present case,
a complex scalar (or equivalently $2$-vector) auxiliary field
$m\rt$ is appropriate. The magnitude $|m|$ has the interpretation of
the {\it distance} to the nearest vortex and the phase of $m$ specifies
the relative positions of interfaces. The nonlinear transformation relating
the order parameter field to this auxiliary field
 $\psi\rt =\psi\left(m\rt\right)$
is chosen as the stationary inhomogeneous solution of
\motion\ near an isolated vortex, however with $m$ the coordinate.

We now proceed to evaluate the correlation function by writing
$$ C\rt =<\psi(m_1)\psi^*(m_2)>
        =<\psi_{far}(m_1)\psi^*_{far}(m_2)>+ O(L^{-1}(t))
               \eqn\long  $$
where $m_1\equiv m(\vecr,t)$, $m_2\equiv m(0,t)$. The average is
now over probability distributions of the auxiliary
field $m$. It is essential to note that by extracting out the
far field term in \long\ we have absorbed any non-universal,
potential-dependent contributions into a term of order
$O(L^{-1})$\rlap.\refmark{\fong}
Hence, in the scaling limit $L(t)\rightarrow \infty$
the asymptotic correlation function is independent of the detailed
structure of the potential. Anticipating scaling in this case we write
$$ {\Cal F}_q  = <\psi_{far}(m_1)\psi^*_{far}(m_2)>. \eqno\eq $$
It is easy to show from \farfield\ that
$$ g(\theta_1,\theta_2)\equiv\psi_{far}(m_1)\psi^*_{far}(m_2)=
   e^{2\pi i (k-l)/q}  \eqno\eq $$
for $\theta_1\in\left[\frac{2\pi(k-1)}{q},\frac{2\pi k}{q}\right]$,
 $\theta_2\in \left[\frac{2\pi(l-1)}{q},\frac{2\pi l}{q}\right]$
and $k,l=1,2,..,q$. Here $\theta_1,\theta_2$ are the polar angles of
$m_1$ and $m_2$ respectively.
Now the average over the auxiliary field is of the form
$$ {\Cal F}_q= \int d^2m_1 d^2m_2 \ g(\theta_1,\theta_2)
    \Phi(m_1,m_2)  \eqn\fone $$
with the probability density function
$$ \Phi(m_1,m_2)=\left(\frac{\gamma}{2\pi <m^2>}\right)^2 \exp\left(
    \frac{-\gamma^2(m_1^2+m_2^2-2fm_1m_2^*)}{2<m^2>}\right) \eqno\eq $$
where $\gamma=(1-f^2)^{-1/2}$ and $f\rt=<m\rt m^*(0,t)>/<m^2>$.
By suitably rescaling the arguments, \fone\ can be cast into the form
$$ {\Cal F}_q(f)= \int \frac{d^2x_1d^2x_2}{4\pi^2\gamma^2} g(\theta_1,\theta_2)
    e^{-x_1^2/2-x_2^2/2+x_1x_2f\cos(\theta_1-\theta_2)}.
  \eqn\grand $$
Note that the dependence of ${\Cal F}$ on $\vecr,t$ is implicitly
contained in the correlator $f(\vecr,t)$.
In principle\rlap,\refmark{\gene} this
function should be determined self-consistently from the equation of
motion. Here we shall only need the property that $f$ scales and
has the short-distance expansion: $f=f(\vecr/L(t))=1-\hbox{const}
 \times (r/L)^2 +.. $.

The radial part of \grand\ can be easily evaluated first to give
$$ {\Cal F}_q= \int_0^{2\pi}\int_0^{2\pi}\frac{d\theta_1d\theta_2}{
  4\pi^2\gamma^2}\
   g(\theta_1,\theta_2)Q\left(f\cos(\theta_1-\theta_2)\right)
     \eqno\eq$$
where
$$ Q(\alpha)=\frac{1}{1-\alpha^2}+\frac{\alpha}{(1-\alpha^2)^{3/2}}
   \left(\frac{\pi}{2}+\arcsin\alpha\right).  \eqno\eq $$
It is advantageous to exhaust the symmetry properties of $g$ and
further restrict the limits of integration to obtain more explicitly that
$$\eqalign{
   {\Cal F}_q=\int_0^{2\pi/q}\int_0^{2\pi/q}
  & \frac{d\theta_1d\theta_2}{4\pi^2\gamma^2} \ [
 qQ\left(f\cos(\theta_1-\theta_2)\right)  \cr
  & + 2\sum_{j=1}^{q-1}
  (q-j)\cos\left(\frac{2\pi j}{q}\right)Q\left(f\cos(\theta_1-
    \theta_2-2\pi j/q) \right)  ]. }   \eqn\ftwo $$
Finally we complete the $\theta$
integrations in \ftwo\ and obtain the main result of our paper
$$ {\Cal F}_q[f] = \frac{q}{2\pi^2}\sin^2\left(\frac{\pi}{q}\right)
   \sum_{j=0}^{q-1}\cos\left( \frac{2\pi j}{q}\right)
    (\pi \lda_j+\lda_j^2)   \eqn\main $$
where $\lda_j=\arcsin(f\cos\frac{2\pi j}{q})$.

Equation \main\ shows explicit dependence of the scaling function
upon the degree of degeneracy of the ground states $q$. Several
limits can be scrutinized at once. In the Ising case of $q=2$ one
recovers the well-known result\refmark{\ojk,\gene,\fongtwo}
 for the scaling function
 $$ {\Cal F}_2 =(2/\pi)\arcsin f.  \eqno\eq $$
The case $q=3$ is interesting because
the $3$-state clock model is identical to the standard $3$-state
Potts model. In this case \main\ gives
$$ {\Cal F}_3= \frac{9}{8\pi^2}\left(
     \pi \arcsin f +\pi \arcsin\frac{f}{2}
      +(\arcsin f)^2 - (\arcsin \frac{f}{2})^2 \right) .
  \eqno\eq $$
The scaling function for $q=4$ gives the interesting result
$ {\Cal F}_4=(2/\pi)\arcsin f  $
which is identical to ${\Cal F}_2$. This surprising result
at first sight is, upon reflection not unexpected, since it is
known that the $4$-state clock model is isomorphic to a pair
of non-interacting Ising models\rlap.\Ref\betts{
D.D. Betts, \journal Can. J. Phys. &42&1564(64); M. Suzuki,
 \journal Prog. Theor. Phys. &37&770(67).}
Note that this behavior
occurs due to the accidental symmetry property of the model, therefore
we do not expect this to recur for larger $q$.
While we could similarly regard the $8$-state clock model
as isomorphic to four Ising models, these Ising models are now coupled and
there is no simple relation to the uncoupled Ising type dynamics.
It can be shown that \main\ can be further simplified if even and
odd $q$ are treated separately.
We shall however not dwell on this any further. Let us finally examine the
limit $q\rightarrow \infty$ when we expect the $\infty$-state clock
model to approach the planar rotator or $O(2)$ model. To see this
explicitly, we may simply replace the sum in \main\
by an integral as $q\rightarrow\infty$. Straightforward algebra leads to
$$ {\Cal F}_{\infty}
   =f\int_0^1 \left(\frac{1-z^2}{1-f^2z^2}\right)^{1/2} dz
 = \frac{\pi f}{4} F\left(\frac{1}{2},\frac{1}{2};
   2;f^2\right) \eqn\finf $$
where $F$ is the hypergeometric function. Equation \finf\ is
of the same form as is obtained for the $O(2)$ model\rlap.\refmark{\bray,\fong}

We next proceed to analyze the short-distance behavior of the
scaling function since it gives rise to the anomalous power-law
decay of the scattering form factor $\Phi(k)$ at large wavenumbers.
The limit of short-distances,
i.e., $x=r/L(t)\rightarrow 0$, is associated with the limit
$f\rightarrow 1$. Keeping the leading terms for $f$ near $1$, \main\
can be rewritten in the form
$$ {\Cal F}_q(f)=1-a(q)(1-f^2)^{1/2}-b(q)(1-f) +... \eqn\expand $$
where
$$ a(q)=\frac{q}{\pi}\sin^2\left(\frac{\pi}{q}\right)   \eqno\eq $$
and $$ b(q)= \frac{q}{\pi}\sin^2\left(\frac{\pi}{q}\right)
   \sum_{j=1}^{q-1}\frac{1-2j/q}{\sin(2\pi j/q)}
   +\frac{1}{\pi}\sin\left(\frac{\pi}{q}\right)\left[(q-1)\sin\frac{\pi}{q}
        +\cos\frac{\pi}{q}\right].  \eqn\bexp $$
Recalling that $1-f\sim x^2$ at small $x$, the second term on the
right hand side of \expand\
gives a linear dependence on $x$, for any finite value of $q$.
This can be understood as the consequence of the presence of interfaces
in the system. Alternatively speaking, this term is responsible for
the power-law decay $\Phi (k)\sim k^{-(d+1)}$ at large $k$.
Thus  Porod's law remains
valid for all $q$-state clock models. However, the Porod's tail
gets weaker as the degeneracy $q$ increases. Evidently, as
the limit $q\rightarrow \infty$ is approached, we have
$a(q)\sim q^{-1}\rightarrow 0$. More importantly, notice that
the expansion \expand\ is no longer valid in the limit $q\rightarrow \infty$,
due to the non-interchangeable order of the two limits $\lim_{f\rightarrow 1}$
and $\lim_{q\rightarrow \infty}$. And the coefficient $b(q)$ diverges
$$ b(q) \sim 2\pi \ln q + O(1).  \eqno\eq $$
The logarithmic divergence hints some novel behavior. In fact, in the
limit $q\rightarrow\infty$, we have using \finf\
instead of \bexp, the result
$$ {\Cal F}_{\infty}= 1 + \frac{1}{4}(1-f^2)\ln (1-f^2) +... \eqno\eq $$
which gives a $x^2\ln x$ singularity at small $x$. In Fourier space,
this corresponds to $\Phi(k)\sim k^{-(d+2)}$, the recently
discovered result for the $O(2)$ model\rlap.\refmark{\nigel
-\braytwo,}\Ref\wong{
A.P.Y. Wong, P. Wiltzius, and B. Yurke, \journal Phys. Rev. Lett.
&68&3583(92).}

The methodology we adopted here for the clock model
 could in principle be applied to
the Potts model as well. And it is conceivable that our general
conclusion about the $q$ dependence of the scaling function is correct
in the latter case too.  This raises the need for more accurate
numerical simulations or experiments to further establish
this dependence. Finally, the dynamics for the conserved clock model
is certainly worthy of investigation. Knowing that $L(t)\sim t^{1/3}$ for
$q=2$ while $L(t)\sim t^{1/4}$ for $q=\infty$\rlap,\refmark{\nigel}
there must be crossover behavior as $q$ is varied.

\bigskip
\ack
This work was partially supported by the National Science
Foundation through grant NSF-DMR-91-20719.
F.L. acknowledges partial support from NSF-STC-88-09854
administered through the NSF Science and
Technology Center for Superconductivity.

\endpage
\refout
\end